\documentclass[intlimits,twoside,a4paper]{article}
\usepackage{amsmath,amssymb}
\usepackage{graphicx}
\usepackage{wrapfig}
\usepackage[T2A]{fontenc}
\usepackage[cp1251]{inputenc}

\usepackage[eqsecnum]{cmpj2}

\issue{2013}{16}{1}{13605}

\doinumber{10.5488/CMP.16.13605}

\title[Investigation of nanoporous material]%
{Investigation of nanoporous material under quasi-equilibrium
conditions}

\author[O.V. Yushchenko, T.I. Zhylenko]{O.V. Yushchenko, T.I. Zhylenko}

\address{
Sumy State University, 2 Rimskii-Korsakov St., 40007
Sumy,  Ukraine}

\date{Received October 16, 2012, in final form December 5, 2012}
\begin{document}

\maketitle

\begin{abstract}
Based on the three parametric Lorenz system,  a model was developed  that permits to
describe the behavior of the plasma-condensate system near phase equilibrium in a self-consistent way.
Considering the effect of fluctuations of the growth
surface temperature, the evolution equation and the corresponding
Fokker-Planck equation were obtained. The phase diagram is built which
determines the system parameters corresponding to the regime of the
porous structure formation.


\pacs 64.70.fm, 68.43.Hn, 05.10.Gg
\end{abstract}

\section{\label{sec:level1}Introduction}

\begin{wrapfigure}{i}{0.49\textwidth}
\vspace{-2ex}%
\centerline{\includegraphics[width=0.48\textwidth]{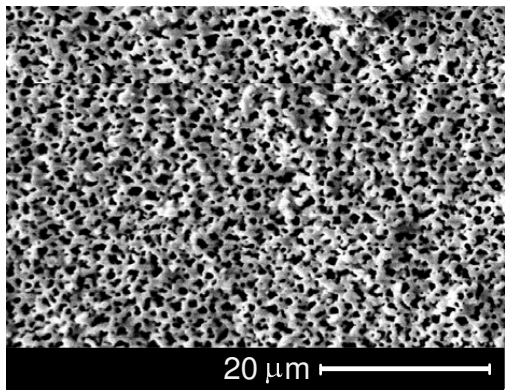}}
\caption{Porous structure of the copper, obtained by spray-deposited material at high discharge
power~\cite{POKK-3, FTT}.} \label{copper}
\vspace{-2ex}%
\end{wrapfigure}

Nowadays, modern nanotechnologies are developed by
using a variety of methods, one of which is the condensation process
in the steady state close to phase equilibrium. This method
makes it possible to obtain various structures of a condensate, fractal
surfaces, porous structures, etc. \cite{POKK-3, FTT}. The
principal feature of this condensation process is the plasma-condensate system being close to phase equilibrium. Consequently, the adsorbed atoms are arranged on the active centers of
crystallization forming structures with different architectures.
In quasi-equilibrium conditions for continuous copper condensation
of about 7 hours, the formation of highly porous structures,
whiskers, and some intermediate structures (fibrous structures with
alternating crystalline and porous parts) was observed. Of particular interest is the structure shown in figure~\ref{copper},
which unlike ordinary single crystals, is realized under unstable
temperature regime.

These structures may be of great practical inte\-rest. For example, they can
be used as a molecular sieve.

However, a question arises regarding the reasons of holding the
plasma-condensate system near phase equilibrium.  Taking into
account a universal nature of the condensation process, we assumed
earlier \cite{FTT,MNT,UFZh,OYBZKP_18_3} that it is caused by
self-organization of a multi-phase plasma-condensate system. From
the physical point of view, the mentioned self-organization is explained
by an increase of the energy of the adsorbed atoms which results
in the temperature increase of the growth surface under the
effect of the plasma within the condensation process. On the
other hand, an increase of the growth surface temperature is
compensated by the desorption flow of the adsorbed atoms, which are
responsible for supersaturation. As a result, within the
framework of synergetic ideology \cite{Haken}, our consideration
is based on the three-parameter Lorenz system \cite{Ol_13,Ol_14}.

The paper is organized as follows. In section~\ref{sec:level2},
the self-organized system that forms the basis of our consideration
is described. Section~\ref{sec:level3} discusses the statistical
analysis of the motion equation. The stationary solution of the
Fokker-Planck equation is considered in section~\ref{sec:level4}.
General conclusions are presented in section~\ref{sec:level5}.

\section{\label{sec:level2}Basic equations}

Considering that a quasi-equilibrium
condensation is provided on the growth surface due to a self-consistent development of the
processes in the plasma volume, we will further use a
three-dimensional (volume) con\-centra\-tion of the condensate $N$
and a two-dimensional (surface) con\-centra\-tion $n\equiv Na$. Here
$a$ is a scale factor, which plays the role of the lattice parameter
and its value will be determined below.

For a given value of the equilibrium con\-centra\-tion $n_{\mathrm{e}}$, the
increasing supersaturation $n-n_{\mathrm{e}}$ is  provided by the diffusion
component defined by the Onsager relation \cite{Lifshic_15,
Landau_16} for the adsorption flow
\begin{equation}
J_{\mathrm{ad}}\equiv D|\pmb{\nabla} N|\simeq \frac{D}{\lambda}(N_{\mathrm{ac}}-N).\label{1}
\end{equation}
It takes into account that the main decrease in the con\-centra\-tion
value takes place near the cathode layer, whose thickness is determined
by the screening length $\lambda$. The latter and the diffusion
coefficient $D$ are given by the equations \cite{Lifshic_15,
Landau_16}
\begin{eqnarray}
\lambda^{2}=\frac{\varepsilon T_{\mathrm{p}}}{4\pi e^{2} N_{\mathrm{i}}}\,, \qquad
D=\frac{\sigma T_{\mathrm{p}}}{e^{2}N_{\mathrm{i}}}\,, \label{2}
\end{eqnarray}
where $\varepsilon$, $\sigma$ are the dielectric permittivity and
the conductivity of the plasma, respectively, $T_{\mathrm{p}}$ is its temperature measured
in the energy units; $e$, $N_{\mathrm{i}}$ are the charge and the total
con\-centra\-tion of the ions of the deposited substance and the inert
gas.

In the second relation (\ref{1}), it is considered that at the upper
boundary layer of the cathode the volume con\-centra\-tion of the
deposited atoms  is reduced to the accumulated value $N_{\mathrm{ac}}$, and
the lower boundary of this layer presents the growth surface, near
which the con\-centra\-tion of atoms is $N$.

A decrease of supersaturation $n-n_{\mathrm{e}}$ is ensured by the
desorption flow $\textbf{J}$, which is directed up from the growth
surface, so that $J<0$, while the value of the adsorption flow
$J_{\mathrm{ad}}>0$. In case there is no condensate (when all the adsorbed
atoms have evaporated from the substrate), the  condition $J=-J_{\mathrm{ad}}$ is performed for the desorption component. Here, the accumulated flow
$J_{\mathrm{ad}}$ is defined by the equation (\ref{1}), where $N=N_{\mathrm{e}}$. The
diffusion changes of the concentration $N$ of the deposited atoms are presented by the continuity equation $\dot{n}/a+{\pmb{\nabla}}{\textbf{J}}_{\mathrm{ad}}=0$.
Here, the point over $n$ denotes the differentiation with respect to time and
the source effect is given by the estimate
 \begin{equation}
|{\pmb{\nabla}} \textbf{J}_{\mathrm{ad}}|\simeq \frac{J_{\mathrm{ad}}}{\lambda} \simeq
\frac{(D/\lambda^{2})(n-n_{\mathrm{e}})}{a}\,.\label{3}
\end{equation}
Thus, the diffusion dissipation of con\-centra\-tion is
expressed by the equation $\dot{n}\simeq(D/\lambda^{2})(n-n_{\mathrm{e}})/a$.

On the other hand, the velocity of desorption  of atoms
$\int_{v}\dot{N}\rd v$ in volume $v$, based on the growth surface $s$,
is as follows:
\begin{equation}
\int_{\nu} \dot{N}\rd v=-\int_{\nu} (\pmb{\nabla} \textbf{J})\rd v =
-\int_{\bar{S}} \textbf{J} \rd \textbf{s},\label{4}
\end{equation}
where the first equation takes into account the continuity
condition, while the second equation considers the Gauss theorem. As a result, the
total change of con\-centra\-tion $n=n(t)$ near the growth
surface is described by the equation
\begin{equation}
\dot{n}=\frac{n_{\mathrm{e}}-n}{\tau_{n}} -J.\label{5}
\end{equation}
At the same time, the characteristic relaxation time of the
supersaturation is determined by the equalities
\begin{equation}
\tau_{n} \equiv \frac{\lambda^{2}}{D}=\frac{\varepsilon}{4\pi\sigma}\,,\label{6}
\end{equation}
the second equality being in agreement with the second relation (\ref{2}).

Within the framework of  the synergetic picture \cite{Ol_14}, the
quasi-equilibrium condensation process is caused by the fact, that
along with an increase of the supersaturation $n-n_{\mathrm{e}}$, the condensed
atoms transfer the excess of their energy to the growth surface. As a
result, its temperature $T$ (measured from the ambient temperature)
increases as well. This enhances the evaporation of the deposited atoms due
to an increase of the absolute value of the desorption flow $J<0$, which
compensates the initial supersaturation.

Thus, an appropriate representation of the sequential picture of
quasi-equilibrium condensation process requires a self-consistent
description of the time dependence of the concentration $n(t)$ of adsorbed atoms, the growth surface temperature $T(t)$ and
the desorption flow $J(t)$. According to \cite{Ol_14}, the evolution
equations of these values contain dissipative components and the
terms presenting positive and negative feedbacks, the balance of
which provides a self-organization process. Thereby, in the
equation~(\ref{5}), the first term on the right hand side represents the
dissipation contribution, and the second term presents a linear
relation between the rate of the con\-centra\-tion changes and the
desorption flow.

The evolution equation for the temperature of the growth surface is
presented  in a similar way
\begin{equation}
\tau_{\mathrm{T}}\dot{T}=-T-a_{\mathrm{T}}nJ+\zeta(t), \label{7}
\end{equation}
where $\tau_{\mathrm{T}}$ is a corresponding relaxation time, $a_{\mathrm{T}}>0$ is
the coupling constant. In contrast to the equation~(\ref{5}), it is assumed
that dissipation leads to the relaxation of the growth surface
temperature to the value $T=0$\footnote{It should be noted that the
condition $T=0$ does not correspond to the absolute zero since the
temperature $T$ is measured from the ambient temperature.}. The
second term represents the nonlinear relationship of $\dot T$ with
con\-centra\-tion and flow.  Since the structure shown in
figure~{\ref{copper}}, was obtained at an unstable temperature regime,
the third term on the right hand side of (\ref{7}) is a stochastic
source of temperature changes representing the Ornstein-Uhlenbeck
process\footnote{Investigation of the temperature fluctuation in the
form of white noise was carried out in {\cite{FTT}}.}:
\begin{equation}
\langle\zeta(t)\rangle=0, \qquad
\langle\zeta(t)\zeta(t')\rangle=\frac{I}{\tau_{\zeta}}
\exp\left\{-\frac{|t-t'|}{\tau_{\zeta}}\right\}.
\label{8}\end{equation}
Here, $I$ is the intensity of temperature fluctuations,
$\tau_{\zeta}$ is the time of their correlation.

To ensure self-organization, it is required to
compensate the negative relationship in the expression (\ref{7})
by a positive component in the evolution equation of the flow:
\begin{equation}
\tau_{J}\dot{J}=-(J_{\mathrm{ac}}+J)+a_{J}nT, \label{80}
\end{equation}
where $\tau_{J}$ is a corresponding relaxation time, $J_{\mathrm{ac}}$ is the
accumulation flow, $a_{J}>0$ is a constant of a positive feedback,
allowing the growth of the $\dot J$ due to the mutual effect of
the concentration of the adsorbed atoms and the growth surface
temperature.

Thus, equations (\ref{5}), (\ref{7}), (\ref{80}) present a synergetic
system, where the super\-sa\-tu\-ra\-tion $n-n_{\mathrm{e}}$ is reduced to
the order parameter, the temperature $T$ of the growth surface --
to the conjugate field, and the desorption flow $J$ -- to the
control parameter \cite{Ol_14}. As a result, the task is to
investigate the possible stationary regimes in a stochastic
plasma-condensate system, in particular, to consider  the regime of
the formation of porous structures.

The most simple investigation of the system
(\ref{5}), (\ref{7}), (\ref{80}) is possible within a dimensionless
form using the characteristic scales for the time $t$, the
con\-centra\-tion $n$, the temperature of the growth surface $T$,
the flow $J$, and for the intensity of the temperature fluctuations
$I$:
\begin{equation}t_{\mathrm{s}}\equiv \tau_{n},\qquad   n_{\mathrm{s}}\equiv a^{-2}, \qquad  T_{\mathrm{s}}\equiv
\varepsilon, \qquad  J_{\mathrm{s}}\equiv \tau^{-1}_{n}a^{-2},\qquad   I_{\mathrm{s}}\equiv
\tau^{-1}_{n}a_J^{-2}, \label{9}\end{equation} where the
above-mentioned length  $a=(a_{\mathrm{T}}a_{J})^{1/4}$ and energy
$\varepsilon=(\tau_{n}a_{J})^{-1}$ were used.

Thus, the dimensionless system of equations describing the
fluctuational transition in a plasma-con\-densate system takes the
form
\begin{eqnarray}
\dot{n}&=&-(n-n_{\mathrm{e}})-J, \nonumber\\
\epsilon\dot{T}&=&-T-n J+\zeta(t), \nonumber\\
\sigma\dot{J}&=&-(J_{\mathrm{ac}}+J)+n T, \label{10}
\end{eqnarray}
where we introduced the relations for the relaxation times
\begin{equation}
\epsilon=\frac{\tau_{\mathrm{T}}}{\tau_{n}}\,,\qquad
\sigma=\frac{\tau_{J}}{\tau_{n}}\,.\label{11}\end{equation}

\section{\label{sec:level3} Statistical analysis }

While this system has no analytical solution, we will use the
approximation $\tau_{n}\simeq \tau_{J}\gg \tau_{\mathrm{T}}$, which means
that the temperature varies most rapidly. This situation is
realized in the experiment very rarely,  but  the structure,
presented in figure~\ref{copper}, is obtained  exactly under unstable
temperature regime (unstable cooling).

Then, on the left hand side of the equation (\ref{80}), we can assume
$\epsilon\dot{T}\simeq 0$, and the conjugate field is expressed by
the equation $T=-nJ+\zeta(t)$.

After some simple mathematical operations \cite{Ol_14, OK-18_1,
Kharch} the system (\ref{10}) reduces to the evolution equation
having a canonical form of the nonlinear stochastic Van der Pol
oscillator \cite{sinchroniz}
\begin{equation}\sigma \ddot{n}+\gamma(n)\dot{n}=f(n)+g(n)\zeta(t).\label{14_1}\end{equation}
Here, the friction coefficient $\gamma(n)$, the force $f(n)$ and the
noise amplitude $g(n)$  are presented by the equations
\begin{eqnarray}
\gamma(n)&=&1+\sigma+n^2,\nonumber\\
f(n)&=& J_{\mathrm{ac}}-(n-n_{\mathrm{e}})(1+n^2),\nonumber\\
g(n)&=& n.\label{gfg}
\end{eqnarray}

Then, the task is to find a distribution  function of the system in
the phase space formed by the con\-centra\-tion $n$ and the rate of its
change $p=\sigma \dot n$ depending on time  $t$.

To this end, the Euler equation (\ref{14_1}) is conveniently
represented by the Hamilton formalism
\begin{eqnarray}
\dot n&=&\sigma^{-1}p,\nonumber\\
\dot p&=&-\sigma^{-1}\gamma(n)p+f(n)+g(n)\zeta(t).\label{Ham}
\end{eqnarray}
Thus, the  above-mentioned probability density $P(n,p,t)$ is reduced
to the distribution function $\rho(n,p,t)$ for the solutions of the
system (\ref{Ham}):
\begin{equation}
P(n,p,t)=\langle\rho(n,p,t)\rangle_{\zeta}\,,\label{49}
\end{equation}
where $\langle\ldots\rangle_{\zeta}$  means the averaging over noise
$\zeta$.

We will proceed from the continuity equation
\begin{equation}
\frac{\partial }{\partial
t}\rho(n,p,t)+\left\{\frac{\partial}{\partial n}\left[\dot n
\rho(n,p,t)\right]+\frac{\partial}{\partial p}\left[\dot p
\rho(n,p,t)\right] \right\}=0.\label{50}
\end{equation}

Further, the substitution of the equalities (\ref{Ham}) leads to the
Liouville equation
\begin{equation}
\left[\frac{\partial}{\partial{t}}+\hat{\mathcal{L}}(n,p)\right]\rho{(n,p,t)}=
-g(n)\zeta(t)\frac{\partial}{\partial p}\rho{(n,p,t)},\label{51}
\end{equation}
where the operator
\begin{equation}
\hat{\mathcal{L}}(n,p)=\frac{p}{\sigma}\frac{\partial}{\partial n}+
\frac{\partial}{\partial
p}\left[f(n)-\frac{\gamma(n)}{\sigma}\right].\label{52}
\end{equation}

Turning to the interaction representation \cite{Shapiro}
\begin{equation}
\varrho(n,p,t)=\re^{\hat{\mathcal{L}}(n,p)t}\rho(n,p,t),\label{53}
\end{equation}
the equation~(\ref{51}) takes the form
\begin{equation}
\frac{\partial \varrho(n,p,t)}{\partial
t}=-\re^{\hat{\mathcal{L}}(n,p)t}g(n)\zeta(t)\frac{\partial}{\partial
p}\re^{-\hat{\mathcal{L}}(n,p)t}
\varrho(n,p,t)\equiv\varepsilon\mathcal{R}(n,p,t)\varrho(n,p,t),\label{54}
\end{equation}
where $\varepsilon$ is a dimensionless small parameter
\cite{Shapiro}. Then, using the cumulant expansion method
\cite{VanK}, one can obtain the kinetic equation\footnote{It takes
into account that the time derivatives in equations~(\ref{51}), (\ref{54})
were treated according to the Stratonovich rule.}
\begin{equation}
\frac{\partial \varrho(n,p,t)}{\partial t}=\varepsilon^2
\int_0^t\left\langle\mathcal{R}(n,p,t)\mathcal{R}(n,p,t')
\right\rangle\langle\varrho(n,p,t')\rangle
\rd t',\label{55}
\end{equation}
neglecting the terms of $\varepsilon^3$ order \cite{Ol_14}.

Since a physical time $t$ is usually much longer than the noise
correlation time $\tau_\zeta$, the upper limit of the integration
can be set equal to infinity. Then, returning from the interaction
presentation  to the original presentation, for the distribution
function (\ref{49}) we obtain
\begin{equation}
\left[\frac{\partial}{\partial{t}}+\hat{\mathcal{L}}(n,p)\right]P(n,p,t)=
\varepsilon^{-2}\hat{\mathcal{N}}P(n,p,t). \label{56}
\end{equation}
Here, $\hat{\mathcal N}$ is a scattering operator, which is given by
the expression
\begin{equation}
\hat{\mathcal N}= \left[M_0(t)-\gamma(n)M_1(t)\right]g^2(n)
\frac{\partial^2}{\partial p^2}+\varepsilon M_1(t)g^2(n)
\left[-\frac{1}{g(n)}\frac{\partial g(n)}{\partial
n}\left(\frac{\partial}{\partial p}+p \frac{\partial^2}{\partial
p^2}  \right)+\frac{\partial^2}{\partial n\partial p}
\right]+\mathcal{O}(\varepsilon^2),\label{57}
\end{equation}
where $M_0(t)$ and $M_1(t)$ are the moments of the correlation
function (\ref{8})
\begin{equation}
M_i(t)=\frac{1}{i!}\int^{\infty}_{0}t^{i}\langle\zeta(t)\zeta(0)\rangle
\rd t.\label{17}
\end{equation}
From equation (\ref{17}) one can obtain
\begin{equation}
M_0(t)=I, \qquad M_1(t)= I\tau_\zeta\,. \label{17_1}
\end{equation}

Since, for this task, a complete distribution function $P(n,p,t)$
has a lower practical interest than its integral
\begin{equation}
\mathcal P(n,t)=\int P(n,p,t)\rd p,\label{19_1}
\end{equation}
it makes sense to consider the moments of the initial distribution
function
\begin{equation}
\mathcal P_i(n,t)=\int p^i P(n,p,t)\rd p.\label{19}
\end{equation}
Then, the zero moment is reduced to the required integral (\ref{19_1}).

Multiplying equation~(\ref{56}) by $p^i$ and integrating over all $p$, we
arrive at the relation that can be written as a Fokker-Planck
equation presented  in the Kramers-Moyal form   \cite{Risken}
\begin{equation}
\frac{\partial \mathcal
P(n,t)}{\partial{t}}=-\frac{\partial}{\partial{n}}\left[D_1(n)\mathcal
P(n,t)\right]+ \frac{\partial^{2}}{\partial n^2}\left[D_2(n)\mathcal
P(n,t)\right],\label{20}
\end{equation}
where the drift coefficient
\begin{equation}
D_1(n)=\frac{1}{\gamma(n)}\left[f(n)-M_0(t)\frac{g^2(n)}{\gamma^{2}(n)}\frac{\partial
\gamma(n)}{\partial n}+M_1(t)g(n)\frac{\partial g(n)}{\partial n}
\right]\label{21}
\end{equation}
and the diffusion coefficient
\begin{equation}
D_2(n)=M_0(t) \frac{g^2(n)}{\gamma^{2}(n)}\label{22}
\end{equation}
are presented by the functions (\ref{gfg}).

\section{\label{sec:level4}Stationary solution}

A stationary solution of the Fokker-Planck  equation \cite{Risken}
yields a stationary distribution \cite{Ol_14, OK-18_1}
\begin{equation}
\mathcal{P}(n)=\frac{Z^{-1}}{D_2(n)}{\exp{\int_0^n\frac{D_1(n')}{D_2(n')}\rd n'}},\label{42}
\end{equation}
where the partition function $Z$ is presented by the equation
\begin{equation}
Z={\int_0^\infty\frac{\rd n}{D_2(n)}\exp{\int_0^n\frac{D_1(n')}{D_2(n')}\rd n'}}.
\label{30}
\end{equation}

The extremum condition for the distribution (\ref{42})
\begin{equation}
D_1(n)-\frac{\partial }{\partial n}D_2(n)=0\label{31}
\end{equation}
defines the stationary states of the plasma-condensate system.

\begin{wrapfigure}{o}{0.48\textwidth}
\centerline{\includegraphics[width=0.48\textwidth]{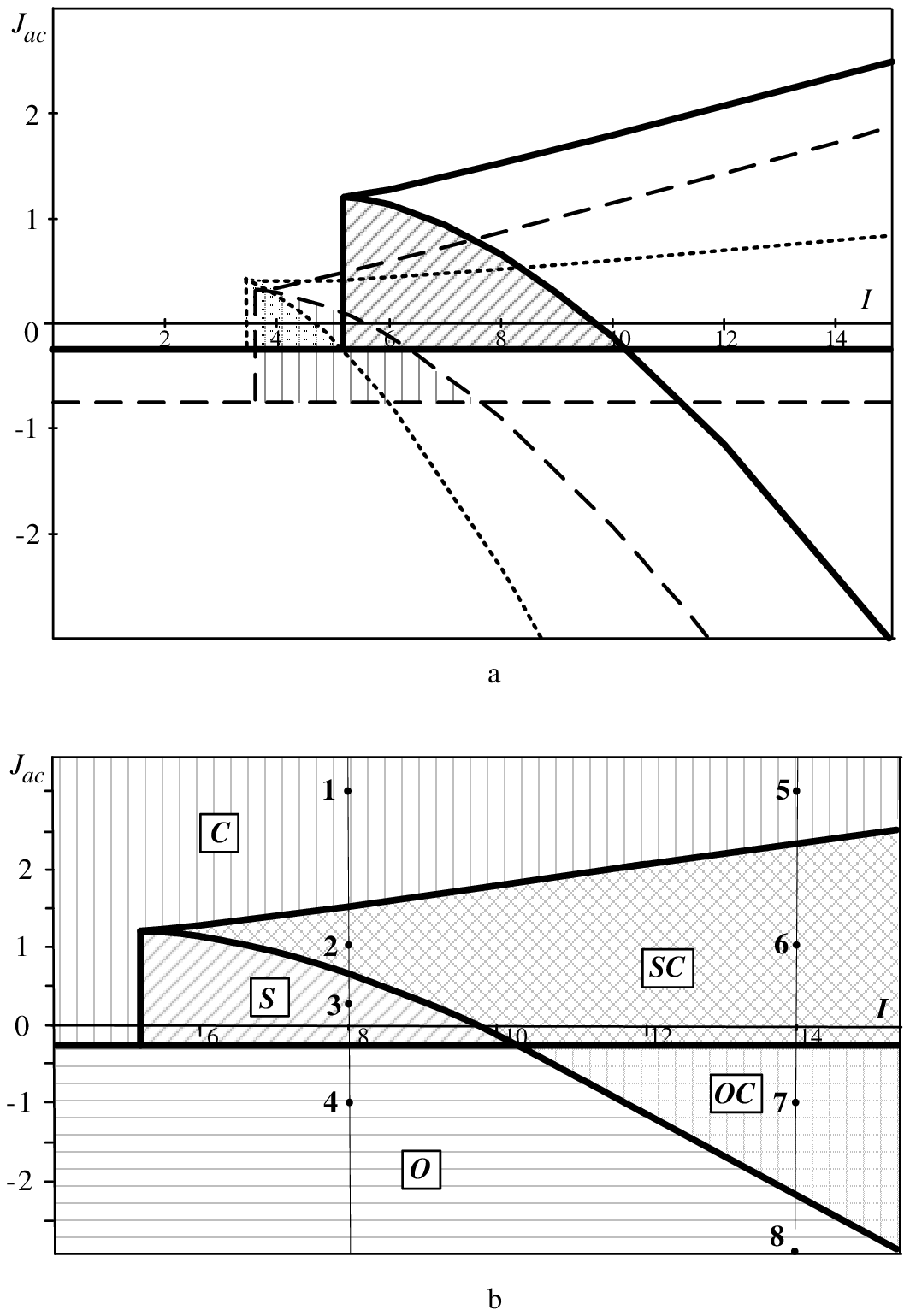}}
\caption{Phase
diagram of the system. The solid line corresponds to
$n_{\mathrm{e}}=0.25$,$\tau_\zeta=0.5$,  the dashed line -- to
$n_{\mathrm{e}}=0.75$,$\tau_\zeta=0.5$, and the dotted line -- to
$n_{\mathrm{e}}=0.25$,$\tau_\zeta=0.75$. The letters indicate the relevant
domains of the phase diagram, and the dots (marked by numbers)
correspond to the parameters at which the stationary concentration
dependence (figure~\ref{St}) is analyzed.} \label{diagr}
\end{wrapfigure}
Substituting  expressions (\ref{21}), (\ref{22}), (\ref{gfg}), and
(\ref{17_1}) into the equation~(\ref{31}) we obtain the equation defining
the stationary concentration dependence
\begin{equation}
J_{\mathrm{ac}}=\frac{2I(1+\sigma)n}{\left[(1+\sigma)+n^{2}\right]^{2}}+(n-n_{\mathrm{e}})(1+n^{2})-I
\tau_{\zeta}n. \label{31_1}
\end{equation}

Then, the condition that restricts the domain of the existence of
the solution $n = 0$ corresponding to the complete evaporation of
the condensate from the growth surface, has the form
\begin{equation}
J_{\mathrm{ac}}=-n_{\mathrm{e}}\,. \label{12}
\end{equation}

The corresponding phase diagram of the system is shown in
figure~\ref{diagr}.

While figure~\ref{diagr}~(a) shows the effect of the system parameters
(the equi\-lib\-rium concentration $n_{\mathrm{e}}$ and the correlation time of
fluctuations $\tau_\zeta$), figure~\ref{diagr}~(b) considers in detail
the domains of the phase diagram. In particular, the domain  $C$
corresponds to the condensation process, the domain $S$ is
characterized by the formation of porous structures, and a complete evaporation of the condensed matter takes place at the domain $O$. The domains,
which are indicated by two letters, meet the coexistence of the
above mentioned regimes.

\begin{figure}[!h]
\vspace{-2ex}%
\centerline{\includegraphics[width=150mm]{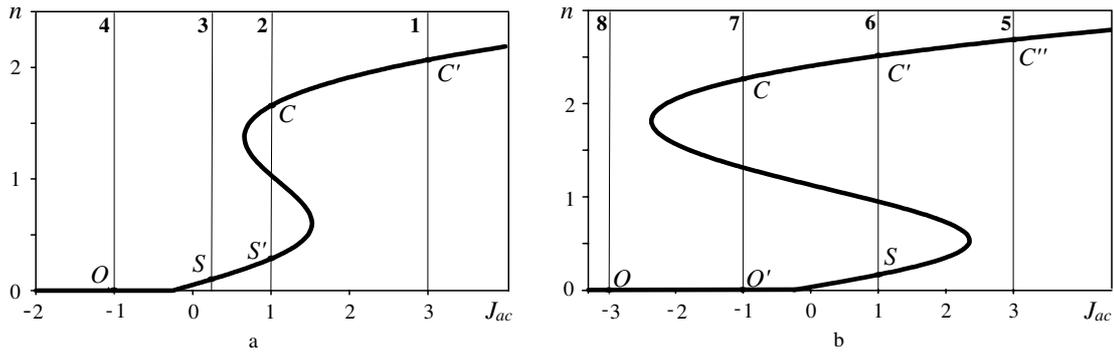}}
\caption{ The
dependence of the stationary concentration $n$ on the accumulated
flow $J_{\mathrm{ac}}$ at $n_{\mathrm{e}}=0.25$, $\tau_\zeta=0.5$, (a) $I=8$, (b)
$I=14$.} \label{St}
\end{figure}

It is more convenient to under\-stand the processes occurring in
each domain considering the example of the stationary concentration
dependence presented in figure~\ref{St}.

Each point on the phase diagram [figure~\ref{diagr}~(b)] corresponds to
the ray in figure~\ref{St}~(a),~(b).

For example, for the ray \textbf{1} [figure~\ref{St}~(a)] only the
condensation pro\-cess is realized. It corresponds to the point
$C'$ characterized by a sufficient stationary concentration $n$.
With a decrease of the accumulated flow (ray \textbf{2}), there is observed
a gradual disassembly of the previously formed condensate. This
situation corresponds to the existence of two steady states with
different concentrations (points $C$ and $S'$).

It should be noted that the additional intersection point of the
main curve and the ray \textbf{2} (which is located between $C$ and
$S'$) applies to the non-physical plot and, therefore, it is not
considered\footnote{This also applies to the similar cases in
figure~\ref{St}~(b).}. Turning to the state of the surface disassembly
(point $S'$), it is worth noting that in this case the usual
evaporation of the upper layer of the condensate does not take place.
First, the atoms which are less connected with the crystallization
centers, are detached from the condensate surface. With a further
decrease of the accumulated flow (ray \textbf{3}), the only  state of
the  surface disassembly remains (point $S$). This is the situation
that characterizes the pattern shown in figure~\ref{copper} and is
of great interest to us. Going to the ray \textbf{4},
the disassembly is replaced by the usual  evaporation process (point
$O$).

Analyzing the relationship (\ref{31_1}) for higher intensity of
fluctuations [figure~\ref{St}~(b)], one can see that some changes occur.
As previously, only condensation process (point
$C''$) is realized for ray \textbf{5}, while ray \textbf{6} is characterized by the
coexistence of disassembly (point $S$) and con\-den\-sa\-tion (point
$C'$) processes. The main difference is found for the ray 7, when
together with the condensation process (point $C$), evaporation
(point $O'$) takes place. At the parameters specified for the ray
\textbf{8}, only evaporation occurs.

\section{\label{sec:level5}Conclusion}

Based on the above analysis, we can conclude that processes
occurring in the plasma-condensate system can be represented within
the system (\ref{5}), (\ref{7}) and (\ref{80})  describing the
self-consistent behavior of concentration, temperature of the
growth surface and desorption flow. Taking into account the fluctuations
of the growth surface temperature with the correlation function
(\ref{8}) makes it possible to describe the most specific
state (disassembly of the surface), when porous nanostructures may
be formed. In addition, as shown in figure~\ref{diagr}~(a), the system
parameters have a significant effect on the domain of the formation of such
structures. With an increase of the correlation time of
fluctuations, this domain significantly decreases and shifts
towards the lower values of the fluctuation intensity, while an
increase in the equilibrium concentration results in a less
significant decrease, as well as causes a shift along two axis (fluctuation
intensity and accumulated flow). As a real
experiment~\cite{POKK-3,FTT}, our theoretical approach has shown
that the state of the surface disassembly  is rarely realized. However,
controlling the parameters of a system, we can reach the regime
under which porous nanostructures are formed.

\section*{Acknowledgements}

The authors are grateful to professor V.I.~Perekrestov for placing the experimental material at their disposal.

\ukrainianpart

\title{Дослідження нанопористих матеріалів за умов квазирівноваги}

\author{О.В. Ющенко, Т.І. Жиленко}
\address{Сумський державний університет, вул. Римського-Корсакова, 2, 40007
Суми, Україна} 

\makeukrtitle

\begin{abstract}
\tolerance=3000%
На основі трипараметричної системи Лоренца була розвинена модель,
що дозволяє самоузгодженим чином описати поведінку системи
плазма-конденсат поблизу фазової рівноваги. Враховуючи вплив
флуктуацій температури ростової поверхні, були знайдені рівняння
еволюції та відповідне рівняння Фоккера-Планка. Побудована фазова
діаграма, на основі якої визначені параметри системи, які
відповідають режимові утворення пористих структур.
\keywords самоорганізація, пересичення, фазова рівновага,
конденсація, фазова діаграма

\end{abstract}

\end{document}